\documentclass[fleqn,usenatbib]{mnras}

\usepackage{newtxtext,newtxmath}

\usepackage[T1]{fontenc}

\DeclareRobustCommand{\VAN}[3]{#2}
\let\VANthebibliography\thebibliography
\def\thebibliography{\DeclareRobustCommand{\VAN}[3]{##3}\VANthebibliography}

\usepackage{graphicx}	
\usepackage{amsmath}	

\title[Two source populations and the knee]{Elemental cosmic ray spectra reveal two populations of Galactic sources \\ and an immediate transition to an extragalactic component after the knee}

\author[T. A. Dzhatdoev \& A. A. Semenov]{
Timur A. Dzhatdoev$^{1}$\thanks{E-mail: timur1606@gmail.com}
and Anatoly A. Semenov$^{2,3,1}$
\\
$^{1}$Institute for Nuclear Research of the Russian Academy of Sciences, 60th October Anniversary Prospect 7a, Moscow 117312, Russia\\
$^{2}$Department of Physics, Federal State Budget Educational Institution of Higher Education \\
M.V. Lomonosov Moscow State University, 1(2), Leninskie Gory, GSP-1, 119991 Moscow, Russia\\
$^{3}$Federal State Budget Educational Institution of Higher Education, M.V. Lomonosov Moscow State University, \\
Skobeltsyn Institute of Nuclear Physics (SINP MSU), 1(2), Leninskie gory, GSP-1, 119991 Moscow, Russia
}

\date{Accepted XXX. Received YYY; in original form ZZZ}

\pubyear{\the\year{}}

\begin{document}
\label{firstpage}
\pagerange{\pageref{firstpage}--\pageref{lastpage}}
\maketitle

\begin{abstract}
The energy spectra for individual elements and/or for groups of elements in cosmic rays (CR) in the energy range between $100 \times Z$~GeV and $10^{3} \times Z$~PeV (where Z is the charge number of the nucleus) have a number of features, including two steepenings (``knees'') with the rigidity-dependent energies $E_{k1} \approx 15 \times Z$~TeV and $E_{k2} \approx 3 \times Z$~PeV and three hardenings (``ankles'') at $E_{a1} \approx 500 \times Z$~GeV; for protons $E_{a2-p} \approx 150$~TeV and $E_{a3-p} \approx 50$~PeV. While the values of $E_{a1}$ for different nuclei are rigidity-dependent, the values of $E_{a2}$ (and probably of $E_{a3}$) are not: $E_{a2-He} \approx 1$~PeV for Helium. The recent advances in precision measurements of the elemental CR spectra in the DAMPE and LHAASO experiments, and, to some extent, in IceTop, KASKADE-Grande, and other experiments, make it possible, for the first time, to clarify the origin of the aforementioned spectral features. We show that the elemental CR spectra are reasonably well described with a sum of three components: 1)~a~low-energy Galactic component with a convex spectral shape reflecting the accelerated particle spectrum in the source; this component peters out after the TeV knee, 2) a high-energy Galactic component including the PeV knee, and 3) an extragalactic component. There is no need for any third, additional component of Galactic cosmic rays in the energy range between 10~PeV and 1~EeV.
\end{abstract}
\begin{keywords}
cosmic rays -- ISM: supernova remnants -- Galaxy: general
\end{keywords}


\section{Introduction}

The all-nuclei cosmic ray (CR) energy spectrum reveals a prominent steepening at $\approx 3-5$~PeV --- the \mbox{so-called} ``knee'' (e.g. \citet{Kulikov1959,Nagano1984,Aglietta1999,Antoni2005,Amenomori2008,Aartsen2013}). The physical origin of the knee is still not certain; most of the existing models are founded on one of the three following effects: 1)~a~break in the spectrum of the accelerated protons/nuclei, often corresponding to the maximum acceleration energy (e.g.~\citet{Peters1961,Stanev1993,Kobayakawa2002}), 2)~a~propagation effect in the Galactic volume \citep{Kulikov1959,Bell1974,Ptuskin1993,Candia2002a,Giacinti2014,Giacinti2015}, or 3) a spectral distortion due to the photopion process, photodisintegration of nuclei or even the Bethe-Heitler process \citep{Karakula1993,Candia2002b,Hu2009}.

By 1995, it was already understood that the Galactic CR spectrum may be composed of at least two distinct components \citep{Stanev1993,Biermann1995}. Some 20--25 years ago, when the number of the available all-nuclei CR spectra from various experiments grew (e.g. \citet{Amenomori1996,Aglietta1999,Fowler2001}) and the statistical uncertainties in the region of the knee became smaller, several phenomenological studies of the knee, as well as of the energy regions below and above the knee, were performed and became popular (e.g. \citet{Hoerandel2003,Hoerandel2004}, hereafter H03, H04). Relatively soon after the publication of H03 and H04, it was recognized that the then available proton CR spectrum measured in various experiments indeed indicated the presence of more than one population of sources \citep{Zatsepin2006} (hereafter Z06); a characteristic signature of a transition between the source populations --- a dip at $\approx 60$~TeV for protons and at a higher energy for heavier nuclei --- is clearly visible in Fig.~1, Fig.~2, and Fig. 4 of this paper.

The proton and Helium CR spectra were measured with the DAMPE detector in the energy ranges of 40~GeV -- 100~TeV \citep{An2019} and 70~GeV -- 80~TeV \citep{Alemanno2021}, respectively. Precision measurements of the proton \citep{Cao2025} and Helium \citep{Cao2026} CR spectra in the knee region reconstructed from observations of extensive air showers (EAS) were recently performed by the LHAASO Collaboration.

The elemental CR spectra in the energy range between $100 \times Z$~GeV and $10^{3} \times Z$~PeV (where $Z$ is the charge number of the nucleus) reveal a number of features, namely: \\
1) a hardening (``ankle'')\footnote{denoted in analogy with a well-known feature in the all-nuclei CR spectrum at $\approx 5$~EeV (e.g. \citet{Lawrence1991,Aab2020a,Abreu2021})} at $E_{a1} \approx 500 \times Z$~GeV; \\
2) a steepening at $E_{k1} \approx 15 \times Z$~TeV (the low energy or TeV knee)\footnote{the spectral structure formed together by the low energy ankle and the TeV knee is sometimes called ``the TeV cosmic-ray bump'' (e.g. \citet{Malkov2021})}; \\
3) a second hardening at $E_{a2-p} \approx 150$~TeV for protons \citep{Varsi2024}, at $E_{a2-He} \approx 1$~PeV for Helium nuclei \citep{Cao2026}, at $E_{a2-Fe} \sim 2$~PeV for Iron nuclei \citep{Apel2013}, and probably somewhere between 1 and 10~PeV for CNO nuclei; \\
4) the ``main'' PeV knee at $E_{k2} \approx 3 \times Z$~PeV; \\
5) another hardening at $E_{a3-p} \approx 50$~PeV for protons; \\
6) in addition to that, the spectrum of Iron nuclei (and possibly of CNO and of the Silicon group) reveals a signature of the spallation process on the interstellar matter and/or inside the sources below the energy of 50~GeV/nucleon \citep{Aguilar2021a,Adriani2021}.

For the case of CR protons the features in the CR energy spectrum were considered in e.g. \citet{Lipari2020}. The energies of both knees are rigidity-dependent to a reasonable precision\footnote{especially of the first one at $\approx 15 \times Z$~TeV, see \citet{Alemanno2026}}. However, while the same is true for $E_{a1}$, this is not the case for $E_{a2}$: indeed, for the rigidity-dependent case one would expect $E_{a2-He} \approx 300$~TeV, while the measured value is $E_{a2-He} \approx 1$~PeV. The values of $E_{a3}$ may be not rigidity-dependent as well.

In this paper we present an interpretation of the DAMPE, LHAASO, IceTop \citep{Aartsen2019}, and KASKADE-Grande \citep{Apel2013} datasets spanning seven decades on the energy (from 40~GeV to 500~PeV). The low-energy ankle could represent a signature of particle acceleration in blast waves. Indeed, this class of models routinely produces spectra with a convex shape (e.g. \citet{Berezhko1996,Ptuskin2010,Ptuskin2011}); these spectra become harder before the eventual downturn (for a recent discussion see e.g. \citet{Hu2026}). In this case both $E_{a1}$ and $E_{k1}$ are rigidity-dependent unless strong energy losses are present or the probability of photodisintegration of nuclei is significant during the acceleration process. Another possibility was considered in \citet{Aharonian2026} for the proton and Helium CR spectra: in this model, the low-energy ankle is produced by the intersection of the spectra of the first and second Galactic components, while the TeV knee is due to a cutoff in the spectrum of the first Galactic CR component.

The second component forming the PeV knee also provides the origin for the second ankle due to the intersection of its spectrum with the spectrum of the first, lower-energy component. Different chemical compositions of the two components induce some deviations of the values of $E_{a2}$ from the rigidity-dependent trend. However, $E_{k2}$ is rigidity-dependent similarly to $E_{k1}$. Finally, a third component appears at $E_{a3-p} \approx 50$~PeV for protons. As we will see, there are reasons to believe that this component is extragalactic; for the case of the proposed scenario the values of $E_{a3}$ are not rigidity-dependent if the composition of the third, extragalactic component is different from that of the second Galactic component forming the knee.

We briefly discuss the proton spectrum of a single CR source in Sect.~\ref{sect:single}. The proton spectrum for a population of sources with varying maximum energy is considered in Sect.~\ref{sect:population}. In Sect.~\ref{sect:separate} we consider separate fits to the proton, Helium, CNO group, and Iron spectra. A joint fit for all four elemental spectra is presented in Sect.~\ref{sect:grandfit}. Sect.~\ref{sect:discussion} contains a brief discussion. Finally, we conclude in Sect.~\ref{sect:conclusions}.

\section{The proton spectrum of a single CR source}\label{sect:single}

\begin{figure}
\vspace{0.1cm}
\includegraphics[width=\columnwidth]{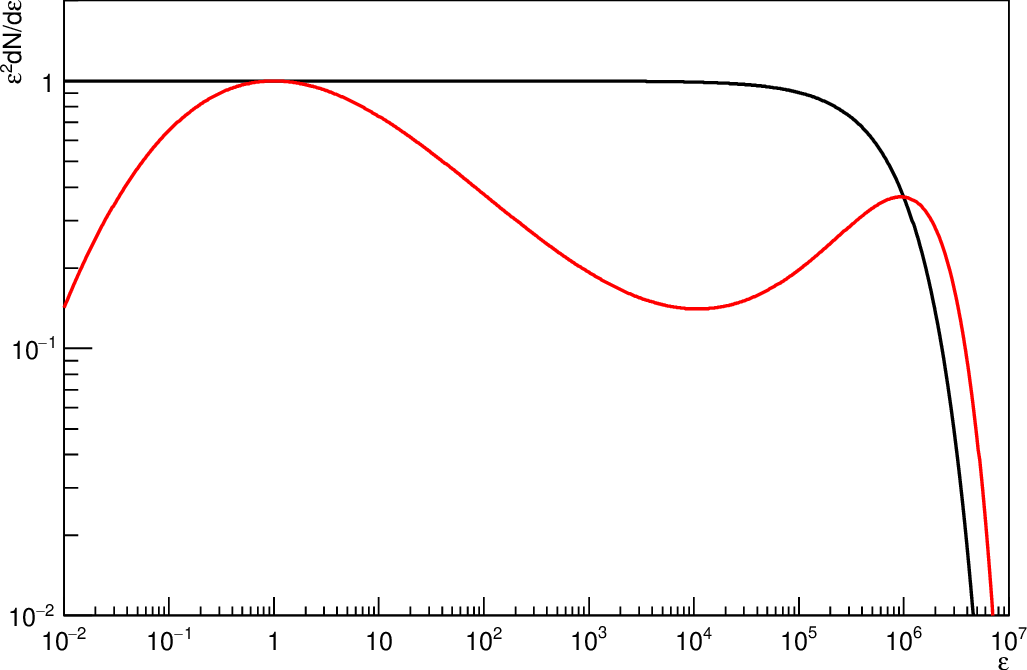}
\caption{Proton spectrum of a single CR source for the EPWL spectral shape (black curve) and the EDLP spectral shape (red curve). \label{fig:OneSource}}
\end{figure}

The simplest commonly used spectral shape for the spectrum of protons accelerated in a Galactic CR source is that of a power-law with an exponential cutoff (EPWL). For the dimensionless kinetic energy $\varepsilon = E/m_{p}$ ($m_{p}$ is the proton mass in energy units) this spectral shape could be expressed as follows:
\begin{equation}
\frac{dN}{d \varepsilon} = K \varepsilon^{-\gamma}exp(-\varepsilon/\varepsilon_{c}), \label{eq-epwl}
\end{equation}
where $K$ is the normalization factor, $\gamma$ is the power-law index, and $\varepsilon_{c}$ is the cutoff energy. The spectral energy distribution (SED = $\varepsilon^{2}dN/d \varepsilon$) for this spectrum is shown in Fig.~\ref{fig:OneSource} (black curve) for $K = 1$, $\gamma = 2$, and $\varepsilon_{c} = 10^{6}$. For supernova remnants (SNRs), a commonly invoked class of Galactic CR sources, a realistic range of values for $\gamma$ is from 1.7--1.8 to 2.2--2.3 and possibly even to 2.4. A low-energy cutoff in the spectrum below $\varepsilon_{c} \sim 10$ is expected from phenomenological considerations; such a cutoff is formally required for $\gamma \ge 2$.

In a realistic astrophysical scenario with the SNRs as the main Galactic CR sources, we expect a number of deviations from the simple EPWL spectral shape, including the following ones: \\
1) the hardening of the spectrum before the cutoff; \\
2) a low-energy downturn in the spectrum; \\
3) the appearance of an additional low-energy component in the spectrum at a relatively late stage of the acceleration process when the velocity of the blast wave drops significantly w.r.t. its maximum value.

The difference in velocity between the shock wave and the surrounding material may be distributed over some distance; furthermore, the diffusion coefficient in the acceleration zone may be energy-dependent. In this case a greater part of the higher-energy particles would be able to move off from the shock front; these particles would probe a greater difference in velocity between the magnetized plasma structures and would therefore undergo a more efficient acceleration, resulting in the hardening of the spectrum before the cutoff. The low-energy downturn in the energy spectrum is sometimes present at $\varepsilon_{r} \sim 1-10$ (see e.g. \citet{Caprioli2012}).

Therefore, we propose a spectral shape more realistic than the EPWL which we call ``the double log-parabola with an exponential cutoff'' (EDLP):
\begin{equation}
\frac{dN}{d \varepsilon} = K \varepsilon^{-\gamma + b \cdot ln( \varepsilon/ \varepsilon_{r})ln( \varepsilon/ \varepsilon_{c})}exp(-\varepsilon/ \varepsilon_{c}), \label{eq-edlp}
\end{equation}
where $b$ is the curvature parameter. The EDLP spectrum is shown in Fig.~\ref{fig:OneSource} as red curve for the case of $K = 1$, $\gamma = 2$, $\varepsilon_{c} = 10^{6}$, $b = 5 \times 10^{-3}$, and $\varepsilon_{r} = 1$. The EDLP spectrum reveals an obvious hardening before the cutoff.

\section{The proton spectrum for \\ a population of sources}\label{sect:population}

\begin{figure}
\vspace{0.1cm}
\includegraphics[width=\columnwidth]{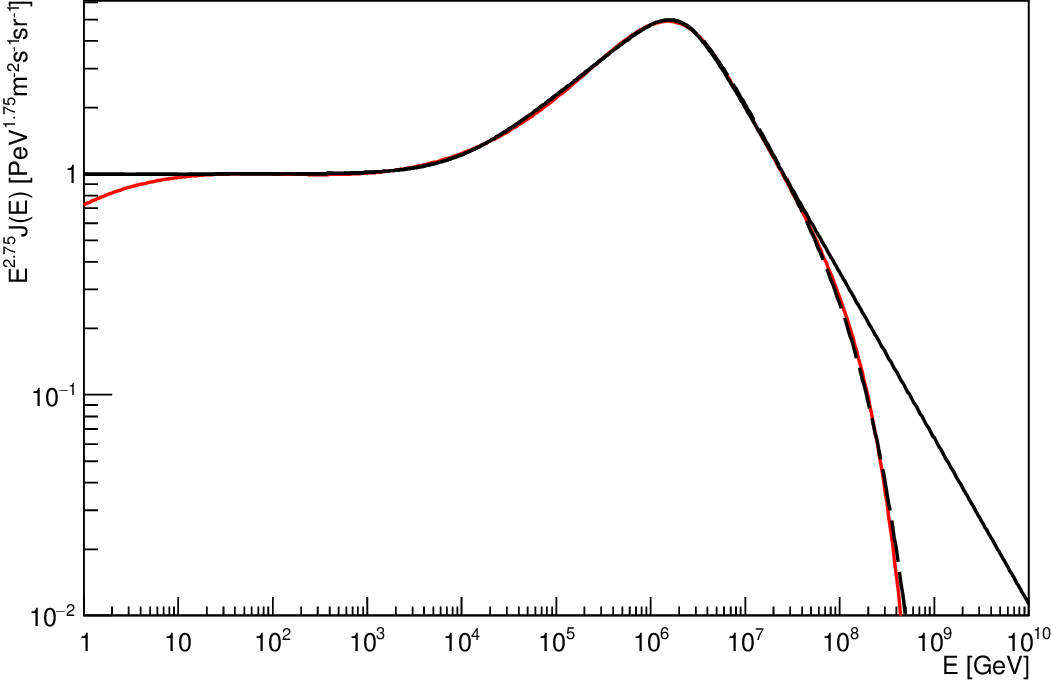}
\caption{Proton spectrum for a population of CR sources defined by eq.~(\ref{eq-pop1}) (red solid curve) and its approximation with eq.~(\ref{eq-pop2}) (black solid curve). Another approximation including an exponential cutoff is shown as black dashed curve. \label{fig:Population}}
\end{figure}

Many different sources with different values of the cutoff energy $E_{c} = m_{p} \varepsilon_{c}$ may contribute to the CR spectrum observed at Earth. Let us assume, for example, that the luminosity of the source is the proxy variable for the maximum proton energy. Then the spectrum of a population of sources $Q(E)$ may be expressed as follows (see e.g \citet{Kachelriess2006,Aloisio2007}):
\begin{equation}
Q(E) = \int \limits_{L_{min}}^{L_{max}} p(L) Q_{s}(E) dL, \label{eq-pop1}
\end{equation}
where $Q_{s}(E)$ is the spectrum contributed by an individual source, $p(L) = p_{0}(L) (L/L_{min})^{- \alpha}$ is the luminosity function, and $L_{min}$ and $L_{max}$ are the minimum and maximum luminosities of the source, respectively. We assume that the cutoff energy is $E_{c}(L) = E_{c0} (L/L_{min})^{\beta}$. We calculate the proton spectrum (see Fig.~\ref{fig:Population}, red curve) for a population of CR sources assuming $\alpha = 2$, $\beta = 1$, $L_{max}/L_{min} = 100$, and $Q_{s}(E)$ is defined by the EDLP function (see eq.(\ref{eq-edlp})) with $\gamma = 2.54$\footnote{the difference between the values of $\gamma$ here and in Sect.~\ref{sect:single} is due to propagation effects that typically result in a steepening of the observable spectrum w.r.t. the spectrum escaping the source}, $b = 3 \times 10^{-3}$, $E_{c0} = 10^{6}$~GeV, $E_{r} = m_{p} \varepsilon_{r} = 1$~GeV. This spectrum (in units of $E^{2.75}J(E)$) was normalized to 1~PeV$^{1.75}$m$^{-2}$s$^{-1}$sr$^{-1}$ at the energy of 100~GeV and then approximated with the following functional form (see Fig.~\ref{fig:Population}, black curve):
\begin{equation}
\begin{split}
J(E) = K \left( \frac{E}{E_{0}} \right)^{-\gamma_{1}} \left [ 1 + \left ( \frac{E}{E_{a}} \right )^{w_{a}} \right ]^{-( \gamma_{2}- \gamma_{1} )/w_{a}} \times \\ \times \left [ 1 + \left ( \frac{E}{E_{k}} \right )^{w_{k}} \right ]^{-( \gamma_{3} - \gamma_{2} )/w_{k}}, \label{eq-pop2}
\end{split}
\end{equation}
with $K = 1.0$, $E_{0} = 100$~GeV, $\gamma_{1} = 2.75$, $\gamma_{2} = 2.40$, $\gamma_{3} = 3.50$, $E_{a} = 10$~TeV, $E_{k} = 2.1$~PeV, $w_{a} = 1.2$, and $w_{k} = 2.6$. At the energy range of 20~GeV--30~PeV the quality of the fit is reasonably good. The ``smoothly broken power law with ankle'' (SBPL-A) functional form of eq.~(\ref{eq-pop2}) coincides with eq. (1) in \citet{Lipari2020}. We note that for the case of $\gamma_{1} = \gamma_{2}$ eq.~(\ref{eq-pop2}) takes the form of a ``pure'' smoothly broken power law (SBPL) functional form and coincides with eq.~(6) in \citet{Ter-Antonyan2000} and eq.~(2) in H03. Finally, another approximation defined by eq.~(\ref{eq-pop2}) with a somewhat different value of $\gamma_{3} = 3.445$ and an additional exponential cutoff $\propto exp(-E/E_{c})$ with $E_{c} = 185$~PeV is shown in Fig.~\ref{fig:Population} as dashed black curve.

\section{Fits to the elemental spectra}\label{sect:separate}

In this and the next Section we set the value of the reference energy \footnote{This auxiliary constant is sometimes called ``the decorrelation energy''} $E_{0} = 100$~TeV. Concerning the absolute energy scale, we choose the measurements of the DAMPE experiment as our benchmark, and we shift all the energy bins in the spectra obtained from EAS observations (namely, in the LHAASO, IceTop, and KASCADE-Grande experiments) by a factor \mbox{$f_{EAS} = 0.93$}; the intensities were corrected accordingly. A similar procedure was utilized in e.g. H03 (Table 3) and \citet{Berezinsky2006} (see their Fig. 11). The LHAASO spectra used in the present paper were obtained assuming the \mbox{EPOS-LHC} hadronic interaction model.

\subsection{A fit to the proton spectrum \label{sect:protons}}

\begin{figure}
\vspace{0.2cm}
\includegraphics[width=\columnwidth]{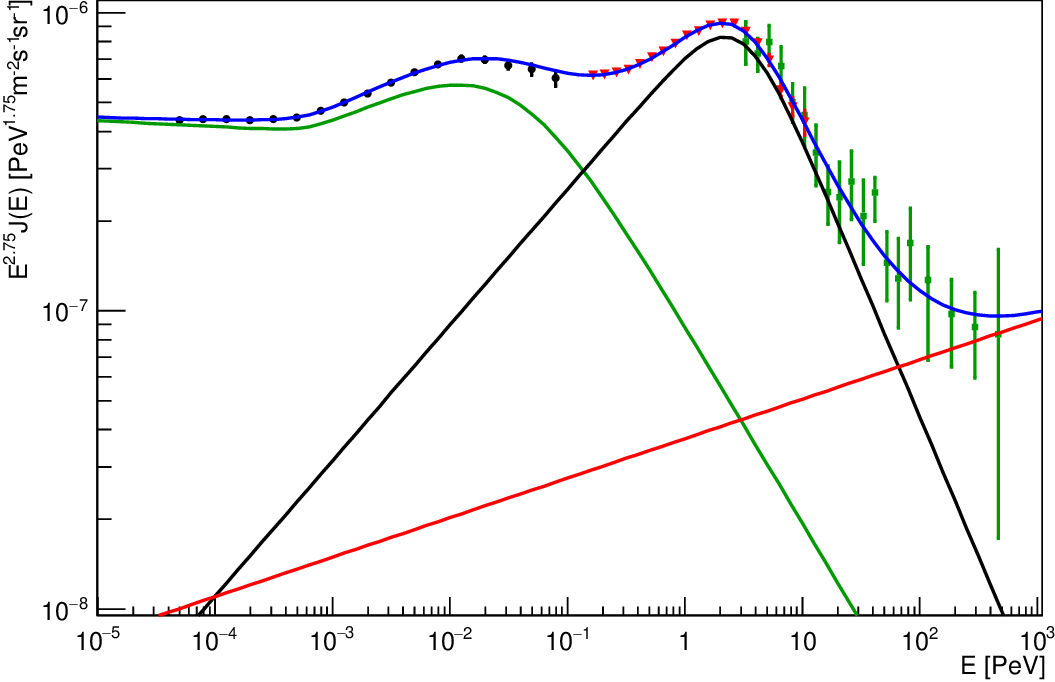}
\caption{Primary proton spectrum measured in various experiments (symbols) together with its fit composed of three components (curves) (see the text for more details). \label{fig:Protons}}
\end{figure}

The proton CR spectrum measured with DAMPE (according to \citet{An2019}) is shown in Fig.~\ref{fig:Protons} as black circles with statistical uncertainties, the LHAASO spectrum is plotted as red triangles; and, finally, the IceTop spectrum is presented as green squares. Some bins in the DAMPE spectrum do not have associated statistical uncertainties (see \citet{An2019}, Table 1). For these, we assumed the statistical uncertainties of $d_{min} = 1$~\%. The last spectral bin of the IceTop spectrum was not included into the fit due to its very large relative statistical uncertainties. We fit the measured proton spectra with the following expression:
\begin{equation}
J(E) = J_{1}(E) + J_{2}(E) + J_{3}(E),
\end{equation}
i.e. a sum of three components, where $J_{1}(E)$ is defined by eq.~(\ref{eq-pop2}) (with free parameters); $J_{2}(E)$ is defined by the same functional form as $J_{1}(E)$ but with different parameters, and $J_{3}(E) = K_{3}(E/E_{0})^{-\gamma_{3}}$ is a simple power-law function. Furthermore, we found that in $J_{2}(E)$ the assumption of $\gamma_{1} = \gamma_{2}$ (hereafter these will be denoted as $\gamma_{21}$ and $\gamma_{22}$, respectively) produces an acceptable fit, and, indeed, we assume $\gamma_{21} = \gamma_{22}$. The fitting was performed with the MINUIT routines \citep{James1975} integrated into the ROOT framework \citep{Brun1997}. The gradient minimization method MIGRAD was adopted. The specific fitting code utilized in the present paper is based on the modified routines used in \citet{Dzhatdoev2020}.

Green curve in Fig.~\ref{fig:Protons} denotes the first (lower-energy) Galactic component, black curve --- the higher-energy Galactic component dominant at the proton knee; red curve denotes the third component, and blue curve is the sum of all these three components. The quality of the overall fit is reasonably good.

We note that another fit to the same dataset obtained with the Markov chain Monte Carlo (MCMC) approach implemented in the emcee code \citep{ForemanMackey2013} yields a similar overall fitting curve (see Appendix~\ref{sect:proton-comparison}). The uncertainties of the parameters were estimated with the MCMC fitting approach. The best-fit parameter values for both fitting methods are presented in Table~\ref{tab:protons} in Appendix~\ref{sect:proton-table}. In particular, for the MIGRAD method we obtained $\gamma_{11} = 2.769$ for the low-energy power-law index of the first Galactic component and $\gamma_{21} = 2.30$; for the MCMC approach the estimates of the same parameters are \mbox{$\gamma_{11} = 2.761_{-0.0096}^{+0.014}$} and $\gamma_{21} = 2.28_{-0.061}^{+0.066}$.

The third (highest-energy) component denoted by red line in Fig.~\ref{fig:Protons} probably has an extragalactic origin. Indeed, the CR flux reveals only very small (if any) observable deviations from isotropy at $E \approx 1$~EeV \citep{Abreu2011,Abreu2012a,Abreu2012b,Abbasi2017,Aab2020b}. In particular, the dipole amplitude at $E = 1-4$~EeV is $r \approx 1$~\% \citep{Abreu2012a,Abreu2012b}, and the 99 \% C.L. upper bounds on the dipole component $d_{\perp}$ below the energy of 8~EeV amount to \mbox{1~\% -- 3~\%} \citep{Aab2020b}.

The available measurements of CR composition, however, reveal the presence of a significant light component at $E \approx 1$~EeV \citep{Aab2014,Abbasi2018,Abbasi2021,Abbasi2026,AbdulHalim2026}. The gyroradius of $\approx 1$~EeV protons in the Galactic magnetic field is comparable to the half-thickness of the Galactic magnetic field disk; therefore, the anisotropy of the Galactic protons at such energies is expected to be significant. Model calculations at $E > 1$~EeV yield $r > 10$~\% for protons \citep{Giacinti2012,Abreu2012a,Abreu2012b}, limiting the contributions of the Galactic protons to the total CR flux to be $f_{p} < 10$~\% at the EeV energy range. Even stronger constrains on the CR EeV Galactic proton content ($f_{p} < 1.3$~\% at 95 \% C.L.) were obtained in \citet{Abbasi2017}.

These results indicate that the light CR component in the EeV energy range has an extragalactic origin. Finally, we note that there is no need for any third, additional component of Galactic cosmic rays in the energy range 10~PeV -- 1 EeV.

\subsection{A fit to the Helium spectrum \label{sect:helium}}

\begin{figure}
\vspace{0.2cm}
\includegraphics[width=\columnwidth]{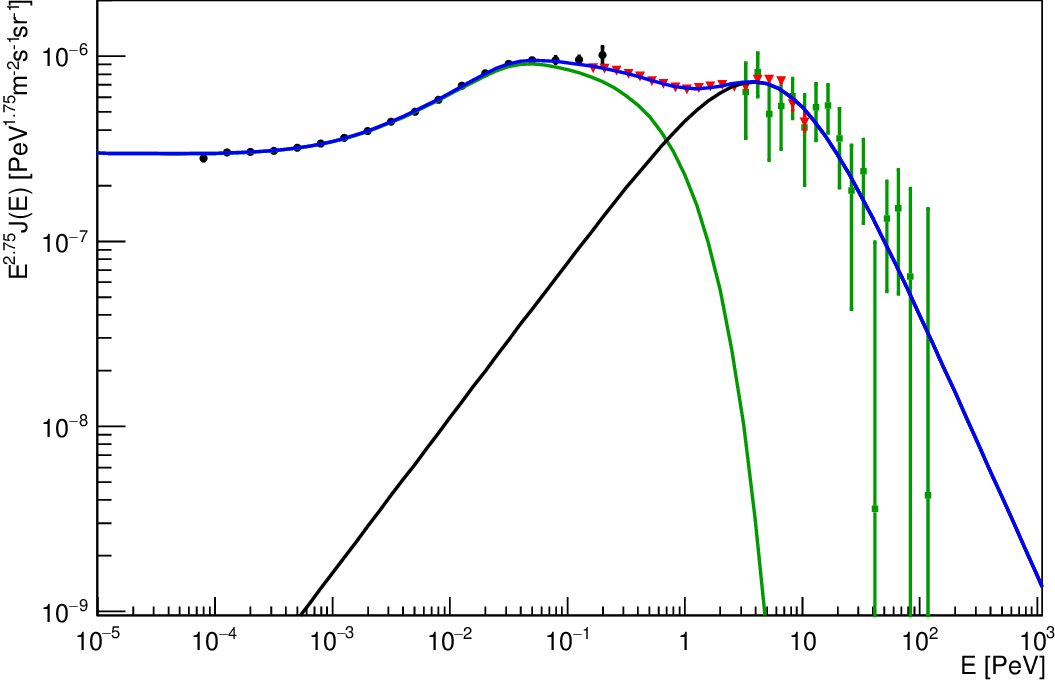}
\caption{Primary Helium spectrum measured in various experiments (symbols) together with its fit composed of two components (curves) (see the text for more details). \label{fig:Helium}}
\end{figure}

A similar fit was obtained for the Helium spectrum (see Fig.~\ref{fig:Helium}). The general meaning of colors in Fig.~\ref{fig:Helium} is the same as in Fig.~\ref{fig:Protons}. In this case the DAMPE spectrum was taken from \citet{Alemanno2026} (hereafter A26). Some bins in the LHAASO spectrum have the statistical uncertainties below 2 \%; for these the statistical uncertainties of 2 \% were assigned. Some bins in the DAMPE spectrum have the statistical uncertainties below $d_{min} = 1$~\%; for these the statistical uncertainties of $d_{min} = 1$~\% were assigned. The last condition will be applied to the Iron and CNO spectra as well (see the next two Subsections). In addition, a 3~\% uncertainty was set to the first bin of the DAMPE spectrum. Only the first 16 bins in the IceTop spectrum were included into the fit. The third (extragalactic) component was excluded from the fit due to the large uncertainties of the IceTop spectrum at $E > 30$~PeV. The value of $E_{a}$ was restricted to be between 800~GeV and 1.5~TeV. Finally, we have introduced an exponential cutoff to the first component at $E_{cm} = 700$~TeV. This is qualitatively similar to the second approximation in Sect.~\ref{sect:population} (see dashed black curve in Fig.~\ref{fig:Population}).

The best-fit parameter values for the Helium spectrum are presented in the second column of Table~\ref{tab:nuclei} in Appendix~\ref{sect:nuclei-table}. We obtained the best-fit value of $\gamma_{11} = 2.759$ (for protons $\gamma_{11} = 2.769$, see Sect.~\ref{sect:protons}). Therefore, the value of $\gamma_{11}$ for protons and Helium could be the same.

\begin{figure}
\vspace{0.2cm}
\includegraphics[width=\columnwidth]{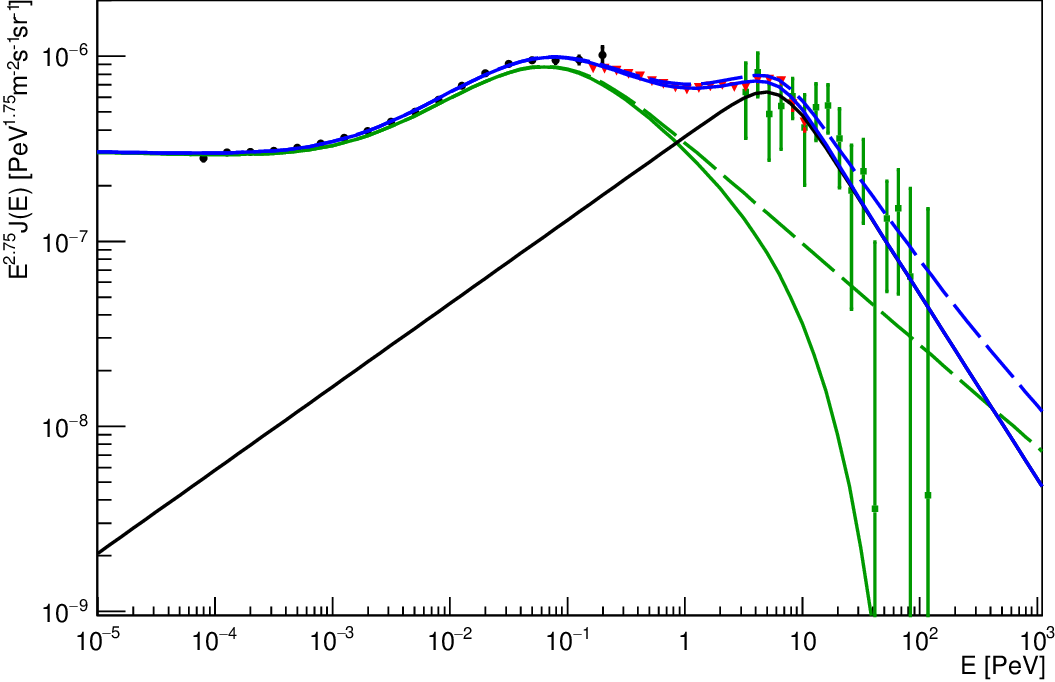}
\caption{Another fit to the primary Helium CR spectrum (see the text for more details). \label{fig:Helium2}}
\end{figure}

A precise measurement of the CR Helium spectrum in the knee energy region is a notoriously difficult problem of astroparticle physics. Even if the statistical uncertainties of this spectrum are small, one could expect significant systematic uncertainties. Concerning the nature of the fitting procedure used in the present paper, relatively small changes in the measured spectrum may have a significant effect on the best-fit values of some parameters. Therefore, it is worthwhile to present another fit to the Helium spectrum which, for example, would allow for a higher value of $E_{cm}$ (see Fig.~\ref{fig:Helium2}). We note that $E_{cm} = 10$~PeV and $\gamma_{11} = 2.766$ for this fit. This fit was obtained after some additional constraints on the parameters were imposed, namely: the value of the power-law index of the first Galactic component after the lower-energy knee $\gamma_{13}$ was restricted to be between 3.3 and 4.0, the value of the power-law index of the second Galactic component before the PeV knee $\gamma_{21}$ --- between 2.3 and 2.7, the value of the ``sharpness'' parameter for the PeV knee $w_{k2}$ --- between 1.0 and 10.0.

In Fig.~\ref{fig:Helium2} we demonstrate an additional option for the fitting spectrum, this time without the exponential cutoff in the second Galactic component (dashed green curve); the total fitting spectrum is denoted as dashed blue curve in this case. The difference of this alternative fitting spectrum from the previous version of the fitting spectrum (solid blue curve) in the energy range where the LHAASO measurements of the CR Helium spectrum were reported is comparable to or lesser than the reported systematic uncertainties of the LHAASO measurement.

\subsection{A fit to the Iron spectrum \label{sect:iron}}

\begin{figure}
\vspace{0.2cm}
\includegraphics[width=\columnwidth]{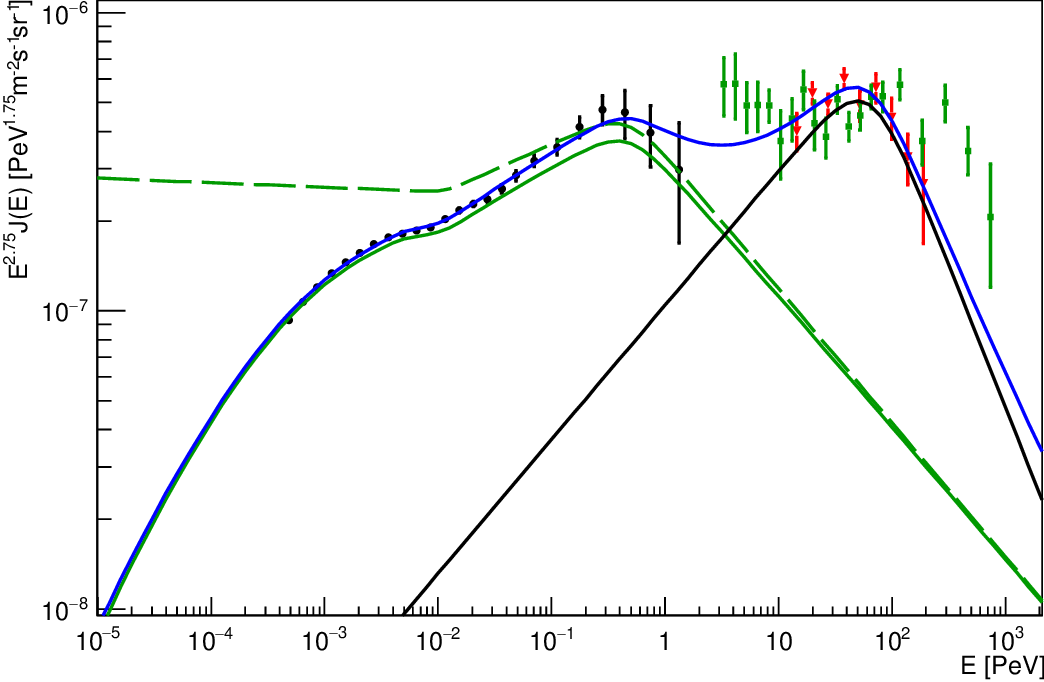}
\caption{Primary Iron spectrum measured in various experiments (symbols) together with its fit composed of two components (curves) (see the text for more details). \label{fig:Iron-Formal}}
\end{figure}

Precision measurements of the Iron spectrum in the knee region are still not available. For the case of the Iron nuclei, we present a fit simular to Figs.~\ref{fig:Protons}--\ref{fig:Helium2} in Fig.~\ref{fig:Iron-Formal}. The general meaning of colors in Fig.~\ref{fig:Iron-Formal} is the same as in Figs.~\ref{fig:Protons}--\ref{fig:Helium2}. The DAMPE spectrum was again taken from A26. As well, the KASCADE-Grande \citep{Apel2013} data were utilized in the fitting procedure. The IceTop spectrum is shown in Fig.~\ref{fig:Iron-Formal} for comparison with the KASCADE-Grande spectrum.

In contrast to the proton and Helium spectra, the spallation process for the Iron nuclei is important at the lower reach of the considered energy range. In order to account for this process, we utilize the total charge changing cross section for Iron presented in \citet{Webber1990} (hereafter W90). The model of the dependence of grammage on the energy $\lambda(E)$ comes from \citet{Aguilar2021a}; it was slightly modified assuming $\lambda(E) \propto R^{-0.45}$ (instead of $\lambda(E) \propto R^{-0.38}$) for the rigidity $R < 192$~GV. We note that \citet{Aguilar2021a} did not include the second (higher-energy) Galactic CR component to their fit; therefore, some differences between their and our models of $\lambda(E)$ could be expected. In the present work this difference is ``absorbed'' into the power-law index of the $\lambda(E)$ dependence at $R < 192$~GV.

The following constraints on the parameters were imposed: $\gamma_{13}$ --- from 3.20 to 3.60, $E_{k1}$ --- from 0.1~PeV to 0.5~PeV, $w_{a}$ --- from 1.0 to 10.0, $w_{k1}$ --- from 2.0 to 3.5, $\gamma_{21}$ --- from 2.30 to 2.50, $\gamma_{22}$ --- from 3.40 to 5.00, $w_{k2}$ --- from 3.0 to 10.0. The best-fit parameter values for the Iron spectrum are presented in the fifth column of Table~\ref{tab:nuclei} in Appendix~\ref{sect:nuclei-table}. We note that $\gamma_{11} = 2.765$ (approximately the same as for the protons and Helium), again, allows one to obtain a reasonable fit. The spectrum of the first (lower-energy) Galactic component without the account of the spallation process is shown in Fig.~\ref{fig:Iron-Formal} as dashed green curve. The systematic uncertainty for both KASKADE-Grande and IceTop is relatively poorly known but significant; for a discussion of the relevant systematics the reader is referred to \citet{Apel2013,Aartsen2019}.

\subsection{A fit to the CNO spectrum \label{sect:iron}}

\begin{figure}
\vspace{0.2cm}
\includegraphics[width=\columnwidth]{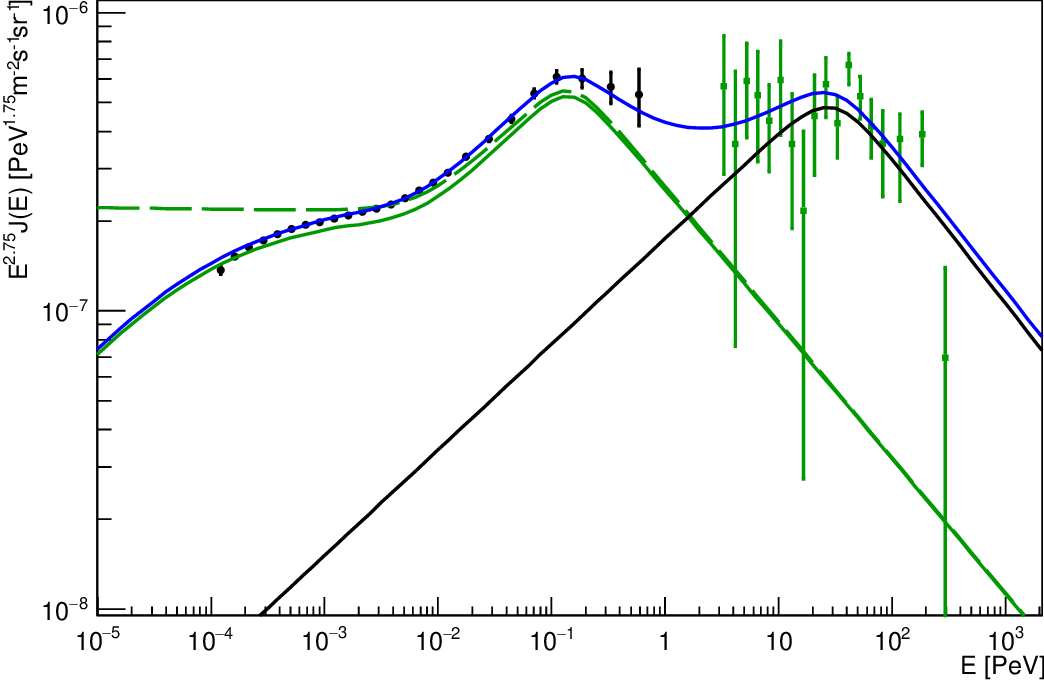}
\caption{Primary CNO spectrum measured in various experiments (symbols) together with its fit composed of two components (curves) (see the text for more details). \label{fig:CNO}}
\end{figure}

Similarly to the case of the Iron spectrum, precision measurements of the CNO nuclei spectrum in the knee region are still not available. The CNO spectrum in the energy range of 2--300~PeV reconstructed in the IceTop experiment (green squares in Fig.~\ref{fig:CNO}) suffers from significant statistical and systematic uncertainties. Only the first 18 bins in this IceTop spectrum were included into the fit. The DAMPE Collaboration reported on their measurement of the Carbon and Oxygen spectra ($J_{C}(E)$ and $J_{O}(E)$, respectively) in A26. We estimate the total (Carbon + Oxygen) intensity as \mbox{$J_{CO}(E)= J_{C}(E) + J_{O}(E)$}; the statistical uncertainty of the total intensity is \mbox{$\Delta J_{CO}(E) = \sqrt{ \Delta J_{C}(E)^{2} + \Delta J_{O}(E)^{2} }$}. The function $E^{2.75} J_{CO}(E)$ is shown in Fig.~\ref{fig:CNO} as black circles. The contribution of the Nitrogen flux to the CNO nuclei spectrum in the GeV--TeV energy domain is modest (see e.g. Sect. 12 of \citet{Aguilar2021b}); it is not considered here.

As for the case of the Iron nuclei, the spallation process was modelled according to W90; here we assume $\lambda(E) \propto R^{-0.50}$ for $R < 192$~GV. Furthermore, we set the following constraints on the parameters: $\gamma_{13}$ --- from 3.20 to 4.20, $E_{a}$ --- from 3.5~TeV to 7.0~TeV, $E_{k1}$ --- from 70~TeV to 150~TeV, $w_{a}$ --- from 1.0 to 10.0, $w_{k1}$ --- from 1.0 to 4.0, $\gamma_{21}$ --- from 2.20 to 2.50, $\gamma_{22}$ --- from 3.20 to 4.50, $E_{k2}$ --- from 15~PeV to 30~PeV, $w_{k2}$ --- from 3.0 to 10.0. The best-fit parameter values for the CNO nuclei spectrum are presented in the fourth column of Table~\ref{tab:nuclei} in Appendix~\ref{sect:nuclei-table}. In this case $\gamma_{11} = 2.754$; this value is approximately the same as for the other considered elemental CR spectra. The fit to the CNO nuclei spectrum is presented in Fig.~\ref{fig:CNO}.

\section{The joint fit to the elemental spectra}\label{sect:grandfit}

\begin{figure*}
\vspace{0.2cm}
\includegraphics[width=\textwidth]{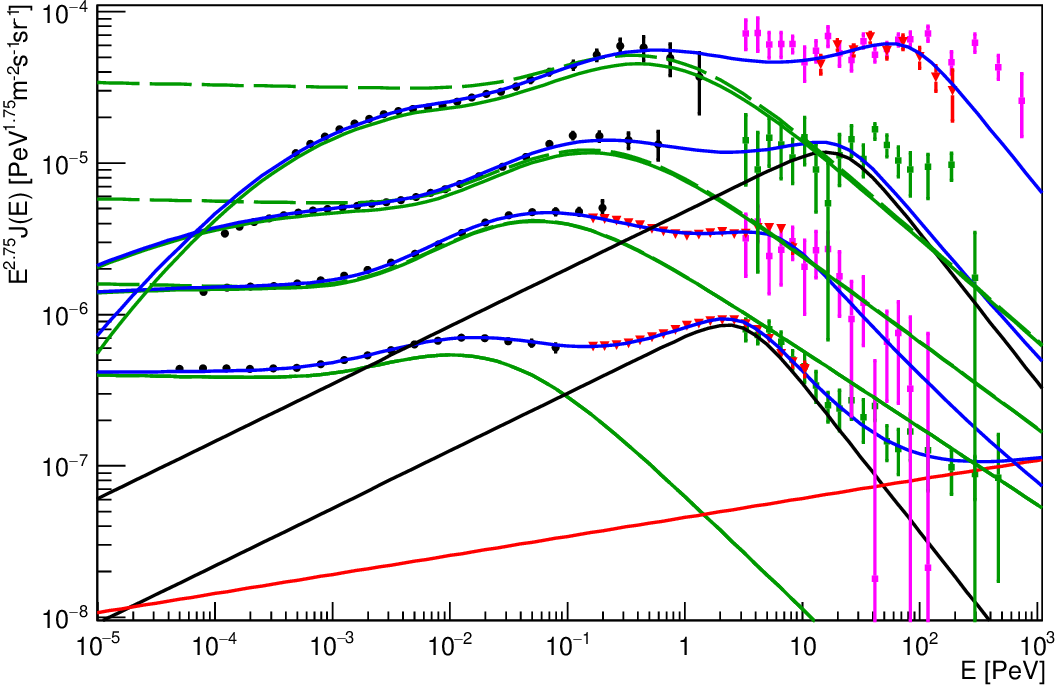}
\caption{The measured elemental CR spectra and the joint fit to the proton, Helium, CNO, and Iron spectra (see the text for more details). \label{fig:Total}}
\end{figure*}

In this Section we present a joint fit for all four elemental spectra together (see Fig.~\ref{fig:Total}). The colors in this figure are the same as in Sect.~\ref{sect:separate} except for the IceTop spectra for the Helium and Iron nuclei; these spectra are shown in magenta. In this Section we assume that $E_{a} = Z E_{a-p}$, $E_{k1} = Z E_{k1-p}$, and $E_{k2} = Z E_{k2-p}$, i.e. that the energies of the low-energy ankle and both knees are strictly rigidity-dependent. We assume $Z = 7$ for the case of the CNO nuclei.

For the Iron and CNO nuclei the spallation process was modelled as before (see Sect.~\ref{sect:separate}); in this Section we assume $\lambda(E) \propto R^{-0.48}$ for $R < 192$~GV. Furthermore, for the Helium nuclei we account for the spallation process as well. We estimate the spallation cross section for the Helium nuclei as $\sigma_{He} \approx (4/56) \sigma_{Fe}$, where $\sigma_{Fe}$ is the spallation cross section for the Iron nuclei.

We denote $\Delta \gamma_{21} = \gamma_{11}-\gamma_{12}$, $\Delta \gamma_{32} = \gamma_{13}-\gamma_{12}$, \mbox{$\Delta \gamma_{2} = \gamma_{22}-\gamma_{21}$}. We set the following constraints on the parameters: $E_{a}$ --- from 400~GeV to 2.0~TeV, $E_{k1}$ --- from 10~TeV to 25~TeV, $w_{a}$ --- from 1.0 to 10.0, $w_{k1}$ --- from 1.0 to 10.0, $K_{3-p}$ --- from 10$^{-5}$~PeV$^{-1}$m$^{-2}$s$^{-1}$sr$^{-1}$ to \mbox{$2 \times 10^{-5}$~PeV$^{-1}$m$^{-2}$s$^{-1}$sr$^{-1}$}, $\gamma_{3}$ --- from 2.4 to 2.7, $\Delta \gamma_{21}$ for the CNO nuclei --- from 0.50 to 0.90. Before the fitting, the intensity of the Iron nuclei measured with the KASCADE-Grande experiment was multiplied by the factor of 0.90 which is smaller than the systematic uncertainty for this spectrum. The total ($\sqrt{\Delta_{stat.}^{2} + \Delta_{syst.}^{2}}$) uncertainty for the CNO spectrum was utilized in the fitting process. Finally, in this Section we set the value of $d_{min} = 2$~\%.

The fit to the elemental spectra obtained with the \mbox{MIGRAD} method is presented in Fig.~\ref{fig:Total}. For better visibility we have multiplied the spectra by a factor of 5$^{n_{v}}$ with $n_{v} = 0, 1, 2, 3$ for the protons, Helium, CNO, and Iron nuclei, respectively. The second (high-energy) Galactic component for the Helium and Iron nuclei is not shown in Fig.~\ref{fig:Total} in order to avoid excessive cluttering of the figure with too many curves. The quality of the fit is reasonably good except for the CNO spectrum. The value of $\Delta \gamma_{21}$ may be somewhat different for the Carbon and Oxygen spectra, inducing some discrepancy between the fit and the measured spectrum around 100~TeV. The spectrum of Carbon contains a significant secondary component from the spallation of the Oxygen and Nitrogen nuclei, and, to some extent, the nuclei heavier than Oxygen (see e.g. \citet{Aguilar2021b,Weng25}). The appearance of this secondary component in the observed spectrum may induce a slight difference between the fit and the measured spectrum in several lowest-energy bins. Finally, a significant discrepancy between the fit and the measured spectrum at $E > 40$~PeV may be due to the systematic uncertainties of the measured CNO spectrum.

The best-fit parameter values for both MIGRAD and MCMC fitting methods are presented in Table~\ref{tab:joint} in Appendix~\ref{sect:joint-table}.  In particular, the values of $\gamma_{11} = 2.765$ for the MIGRAD method and $\gamma_{11} = 2.773_{-0.0065}^{+0.0069}$ were obtained. There is a hint that the values of $\Delta \gamma_{21}$ are indeed different at least for protons and heavier nuclei. The agreement between the overall fitting curves for both methods is reasonably good (see Fig.~\ref{fig:Compare-All} in Appendix~\ref{sect:comparison-all}).

\begin{figure*}
\vspace{0.2cm}
\includegraphics[width=8.5cm]{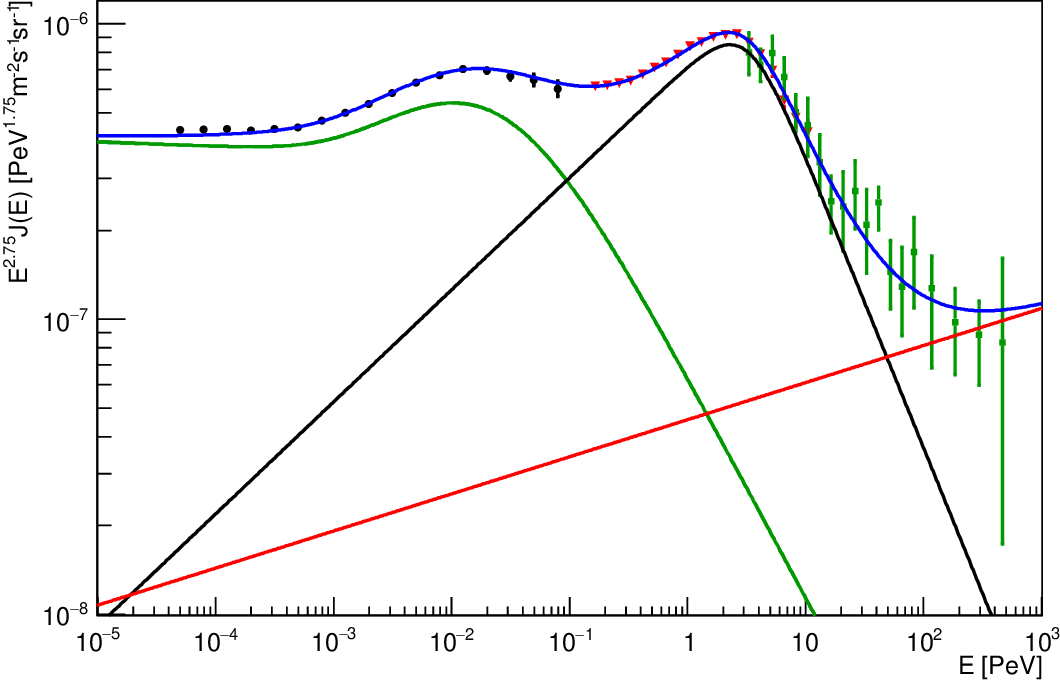}\hspace{0.2cm}\includegraphics[width=8.5cm]{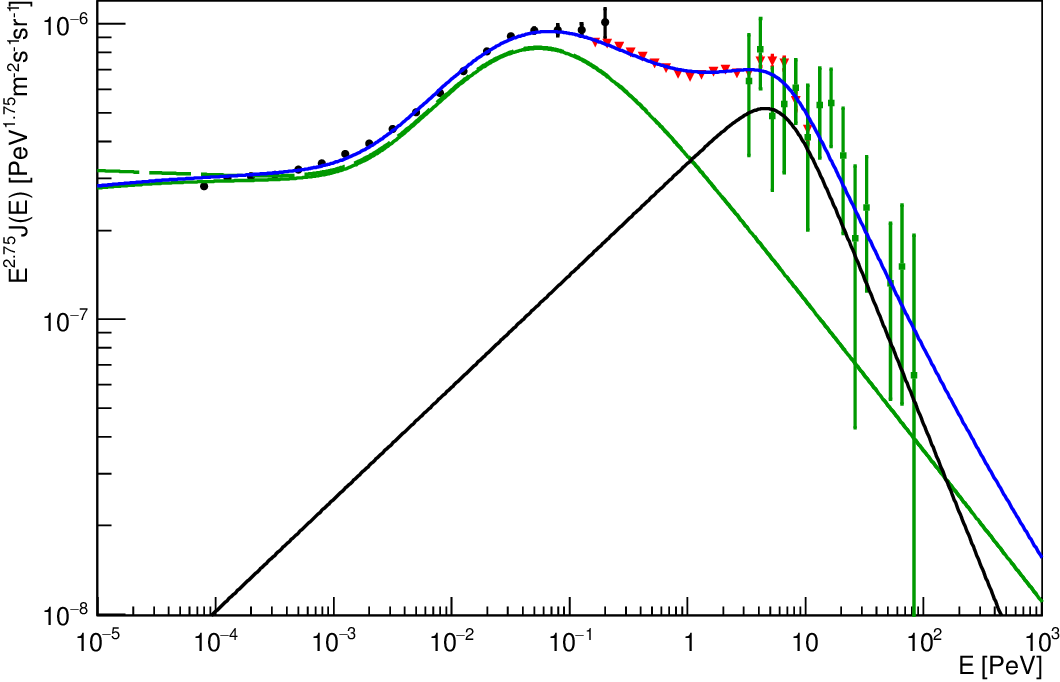}\vspace{0.6cm} \\
\includegraphics[width=8.5cm]{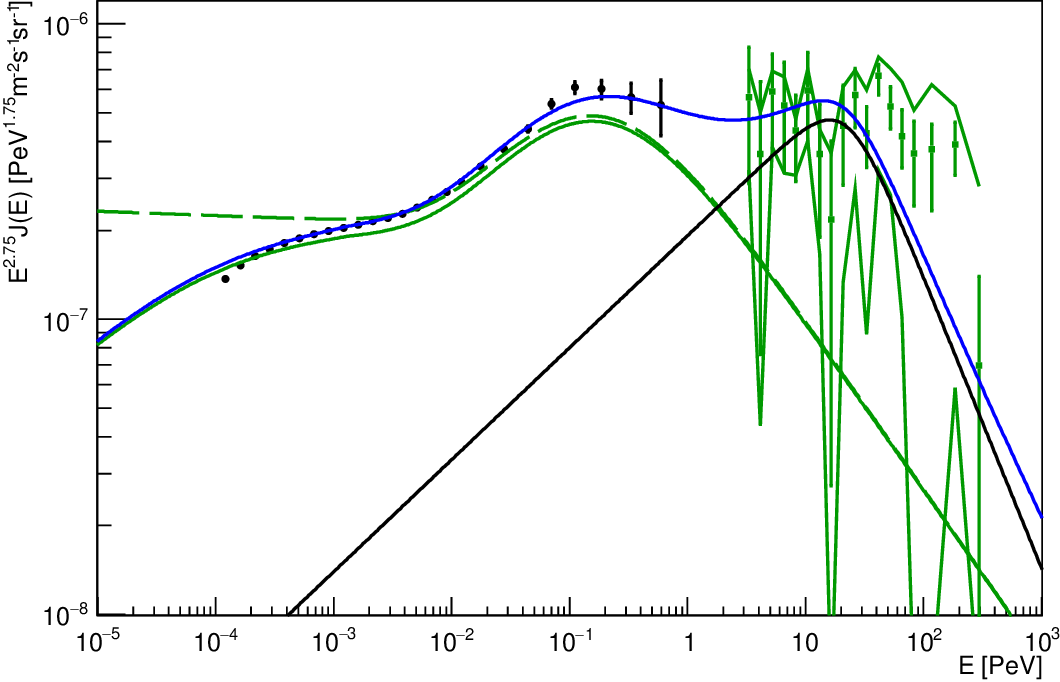}\hspace{0.2cm}\includegraphics[width=8.5cm]{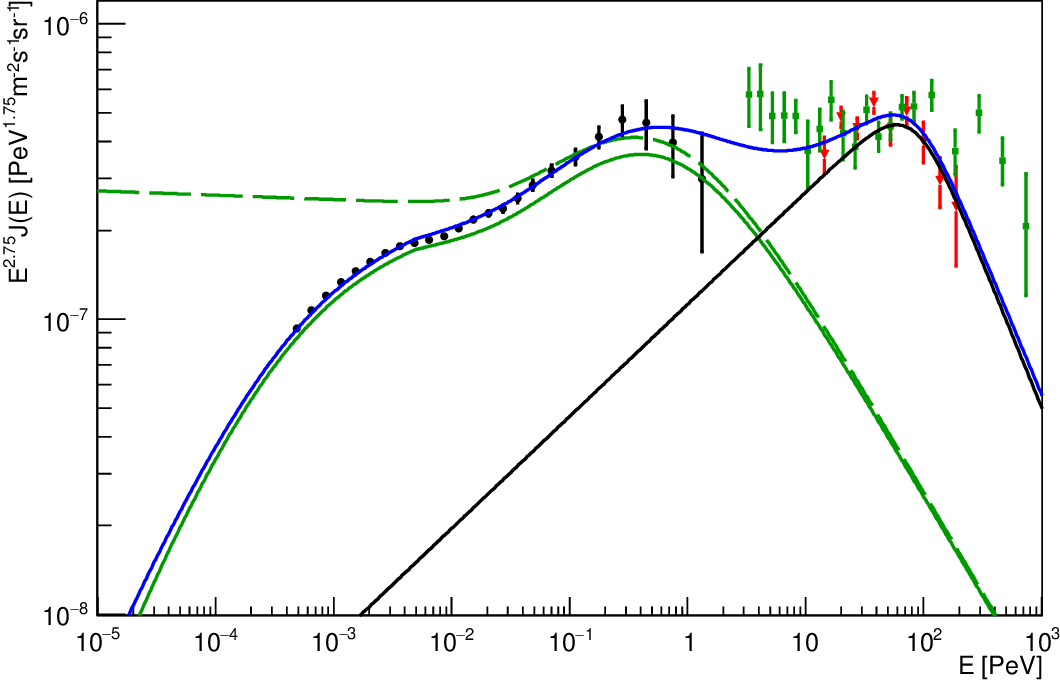}
\caption{The measured elemental CR spectra and the fitting curves from Fig.~\ref{fig:Total} shown separately for the proton (top-left), Helium (top-right), CNO (bottom-left), and Iron (bottom-right) spectra.
\label{fig:Total2}}
\end{figure*}

The same fitting curves as in Fig.~\ref{fig:Total} are plotted in four separate graphs for the proton, Helium, CNO, and Iron spectra in Fig.~\ref{fig:Total2}. For the IceTop CNO spectrum in Fig.~\ref{fig:Total2} (bottom-left) the systematic uncertainties are shown. For the Helium spectrum the two bins with the values of the function $E^{2.75} J_{CO}(E) < 10^{-8}$~PeV$^{1.75}$m$^{-2}$s$^{-1}$sr$^{-1}$ are not shown.

\section{Discussion}\label{sect:discussion}

\subsection{The available CR datasets}

The fits obtained in the present work rely mostly on the DAMPE, LHAASO, IceTop, and KASCADE-Grande datasets. Precision measurements of the elemental CR spectra in the rigidity range between 1~GV and $\approx$100--300~GV were performed with the AMS-02 detector (see e.g. \citet{Aguilar2015a,Aguilar2015b,Aguilar2017,Aguilar2021a,Aguilar2021b}). The AMS-02 and, to some extent, the CALET \citep{Adriani2022,Adriani2023,Adriani2020,Adriani2021} datasets could be utilized in the fitting process together with the other datasets. We leave this study for the future.

The observation of the TeV ``knee'' was reported in \citet{Atkin2018}. The proton and helium CR spectra in the rigidity range 1~GV--1~TV were measured with the PAMELA detector; the shapes of these two elemental spectra were found to be different \citep{Adriani2011}.

\subsection{The convex shape of the spectrum \\ of the first Galactic component}

As discussed in Sect.~\ref{sect:single}, the spectrum leaving the cosmic ray source that defines the injection spectrum into the interstellar medium might have a relatively complex shape. We started this work assuming that the convex shape of the spectrum of the first Galactic component arises in the sources themselves, probably due to various non-linear effects leading to the hardening of the spectrum before the eventual cutoff. However, another explanation is possible. Namely, a ``pedestal'' in the spectrum with the power-law index $\gamma \approx 2.75$ (after accounting for the Galactic propagation effects) could be created by the acceleration in SNRs at a late stage, while an additional ``pile-up'' with a harder spectrum would come from younger SNRs. In this case there are effectively two components within the first Galactic CR component. Of course, the break in the spectrum may be generated by a propagation effect in the Galactic volume (see e.g. \citet{Blasi2012}). As discussed in Sect.~\ref{sect:grandfit}, there is some indication that the difference between the power-law indices before and after the break $\Delta \gamma_{21}$ is less for the proton spectrum than for the spectra of heavier nuclei. However, we do not claim that this difference is statistically significant. We note that the convex shape of the CR spectrum accelerated in SNRs was discussed in \citet{Hillas2005} (Subsect. 4.2 and Fig. 4) based on the model spectra produced in numerical simulations performed in \citet{Voelk2002,Berezhko2003}.

\subsection{On the rigidity-dependent CR spectra}

Ultrarelativistic CR protons and nuclei propagating through the same magnetic and electric fields ``have nearly identical orbits in the phase space $(\vec{r},\vec{R})$'' \citep{Hanusch2019}. In this respect, it is interesting that the values of $\gamma_{11}$ could be the same for all elemental spectra considered in the present work. Likewise, the values of $E_{a}$, $E_{k1}$, $E_{k2}$ could be rigidity-dependent and the values of $\Delta \gamma_{32}$, $\gamma_{2}$, $\Delta \gamma_{2}$, $w_{a}$, $w_{k1}$, $w_{k2}$ could be the same for these elemental spectra. The difference in the observable values of the power-law indices at $E<E_{a}$ might be explained, among other factors, by the onset of the spallation process at these energies.

\subsection{On the nature of the CR sources}

In Sect.~\ref{sect:population}, we obtained an approximation to the spectrum of a population of Galactic cosmic ray sources, assuming a specific luminosity function. As an example, we assumed that the proxy variable for the cutoff energy in the spectrum of the accelerated particles $E_{c}$ was the luminosity $L$ of the source. We note that $E_{c}$ may depend on other parameters as well, such as the properties of the SNR environment or the velocity of the shock wave. Once again, we note that one possible theoretical scenario includes the acceleration of the Galactic CR protons and nuclei in two distinct classes of SNRs, exploding into the interstellar medium and into stellar winds, respectively (see e.g. \citet{Stanev1993}).

Both Galactic components may include some contributions from relatively local sources that have the distance from the source to the observer $D_{s} < 1$~kpc. According to \citet{Erlykin1997}, the PeV knee itself may be formed by a spike created by a single source above the background of other Galactic CR sources. An example of a local source forming the ``the TeV cosmic-ray bump'' was considered in \citet{Malkov2021}. The scenario proposed in the present work allows for some contribution from the local source(s), but this contribution is not strictly necessary.

\subsection{Secondary nuclei}

The measurement of the ratio between the (mostly) secondary and primary CR nuclei (for instance, Boron/Carbon) is a classic tool of cosmic ray physics. Recently, a hardening with $\Delta \gamma = 0.31 \pm 0.05$ in the CR Boron spectrum at 182$\pm$24~GeV/nucleon was observed in the DAMPE experiment \citep{Alemanno2025}. The exact origin of this hardening is not certain. We note that some contribution to the Boron intensity could come from the sources themselves. In particular, the re-acceleration of the secondary nuclei produced in the sources may be important sometimes.

In view of the above-discussed finding that $\Delta \gamma$ may be smaller for protons than for heavier nuclei, we do not exclude the hypothesis that $\Delta \gamma$ in the Boron and Carbon spectra could be the same. We note that the fits obtained in the present work and in \citet{Alemanno2025} are, in some respects, different. In particular, in this work $\Delta \gamma$ denotes the difference in power-law indices before the account of the spallation process. Furthermore, in our work the joint fit obtained for all considered elemental spectra at once (already a difference w.r.t. many other works) includes the second (high-energy) Galactic component as well as the low-energy ``knee''. Therefore, the best-fit values of some parameters are expected to be somewhat different from other studies which adopted different assumptions.

\subsection{Anisotropy}

Another important tool in CR studies is the measurement and interpretation of the amplitude and phase of the anisotropy. A recent compilation of such anisotropy measurements could be found in e.g. A26 (see their Extended Data Fig. 4).

One important variable in anisotropy studies is the ratio between the strengths of the turbulent and regular magnetic field components in the Galaxy \mbox{$K_{tr} = B_{turb}/B_{reg}$}. For $K_{tr} \ll 1$ (almost regular magnetic field) the excess in the observable flux is expected to come from the inner part of the spiral arm of the Galaxy. For isotropic turbulence and $K_{tr} \gg 1$ this excess flux would point towards a concentration of CR sources. The real situation is somewhere between these extremes.

At relatively low rigidities ($R < 300$~TV), both the amplitude and the phase of the anisotropy could be disturbed by the low-scale magnetic fields (see e.g. \citet{Zhang2020}). At $E > 100$~PeV, a transition between the Galactic and extragalactic CR components occurs. At $E \sim 1$~EeV, the amplitude of the anisotropy is lesser than the one expected at the same energy in the framework of the stationary Galactic model with the sources in the Galactic disk. Finally, above the energy of several EeV the amplitude of the anisotropy is expected to rise again.

\subsection{The transition to extragalactic cosmic rays}

As was discussed earlier in Sect.~\ref{sect:protons}, the proton CR component at $E \ge 1$~EeV is, most probably, almost entirely extragalactic. The nature of the transition between the Galactic and extragalactic CR components is a long-standing astrophysical problem
(e.g. \citet{Szabelski2002,Wibig2005,Aloisio2008,Aloisio2012}). We propose a concept of the ``running'' transition between the Galactic and extragalactic cosmic rays at \mbox{$E_{tr} \approx 50 \times Z$~PeV} in the elemental spectra. In this scenario, the Galactic-extragalactic transition is somewhat similar to the transition between the two Galactic components, but the nature of the higher-energy component is different: it is extragalactic (instead of a higher-energy Galactic component).

The deviations from the rigidity dependence of $E_{tr}$ for different elements are due to the difference between the compositions (and possibly between the spectral shapes) of the elemental spectra of the second Galactic and the extragalactic components. The ankle in the all-nuclei CR spectrum at $\approx 5$~EeV could be formed by the intersection of the spectra of the Galactic and extragalactic components. The Galactic component at $E \ge 1$~EeV is dominanted by heavy nuclei; therefore, the anisotropy constraints in this energy range are not violated in the framework of the proposed model. This scenario is appropriate for mixed composition ultra high energy ($E > 1$~EeV) cosmic ray models such as the so-called ``disappointing model'' \citep{Aloisio2011}.

\section{Conclusions}\label{sect:conclusions}

In this paper we present an interpretation for the primary proton, Helium, CNO, and Iron cosmic ray spectra in the energy range from 40~GeV to 500~PeV based on the following general assumptions: \\
1) the low-energy ankle is intrinsic to the first Galactic CR component; \\
2) the observable CR flux includes contributions from different sources with different values of the cutoff energy~$E_{c}$; \\
3) the contribution of the extragalactic component to the observable CR proton spectrum at $E > 100$~PeV is significant.

The Galactic CR spectrum in our scenario is composed of two components: 1) a lower-energy component with a convex spectrum before the cutoff that could be approximated by the ``smoothly broken power law with ankle'' (SBPL-A) functional form, 2) a higher-energy component described by the ``smoothly broken power law'' (SBPL) functional form. The two Galactic components may be produced in supernova remnants (SNRs) of different types and/or exploding into different environments.

Quite remarkably, it is possible to achieve reasonable fits to the elemental CR spectra for the rigidity-dependent positions of the low-energy ankle and both ``knees'' ($E_{a}$, $E_{k1}$, and $E_{k2}$) and the same values of the low-energy power-law index $\gamma_{11}$ for all these spectra. We note that the observable spectra of the Iron and CNO nuclei reveal a significant low-energy cutoff due to the spallation process.

We do not exclude a possible appreciable contribution to the observable CR spectrum from the local sources (e.g. \citet{Erlykin1997,Erlykin2005}). However, a significant contribution from the local sources is not strictly necessary.

In order to obtain a good fit of the proton CR spectrum, one needs to introduce a third component that is probably extragalactic as was discussed in Sect.~\ref{sect:protons}. However, there is no need for an additional third Galactic CR component in the energy range of 10 PeV -- 1 EeV. For the CR protons, the transition between the Galactic and extragalactic components occurs at $E \approx 50$~PeV; for the heavier nuclei --- 
at $E \approx 50 \times Z$~PeV (the ``running'' Galactic-extragalactic  transition); the precise energy of the transition for these nuclei depends on the difference between the compositions and/or the spectral shapes of the elemental spectra of the second Galactic and the extragalactic CR components. In the framework of the scenario adopted in the present paper, the low-energy ``shoulder'' of the ankle in the all-nuclei CR spectrum at $\approx 5$~EeV could be formed by the heavy nuclei of the second Galactic CR component, while its high-energy ``shoulder'' is provided by the extragalactic CR component. Additional studies are required in order to ascertain the physical origin of the EeV ankle in the CR spectrum.

\section*{Acknowledgements}

Helpful discussions with Dr. I.A.~Kudryashov, Ms. A.I. Peryatinskaya, and Prof. G.I.~Rubtsov are gratefully acknowledged. All graphs in the present paper were produced with the ROOT software toolkit \citep{Brun1997}. This work is supported in the framework of the State project ``Science'' by the Ministry of Science and Higher Education of the Russian Federation under the contract 075-15-2024-541.

\section*{Data availability statement}

The data underlying this article will be shared on reasonable request to the corresponding author.

\bibliographystyle{mnras}
\bibliography{Knee-Nuclei}


\appendix

\section{Comparison of fits to the proton spectrum obtained with different methods}\label{sect:proton-comparison}

\begin{figure}
\vspace{0.1cm}
\includegraphics[width=\columnwidth]{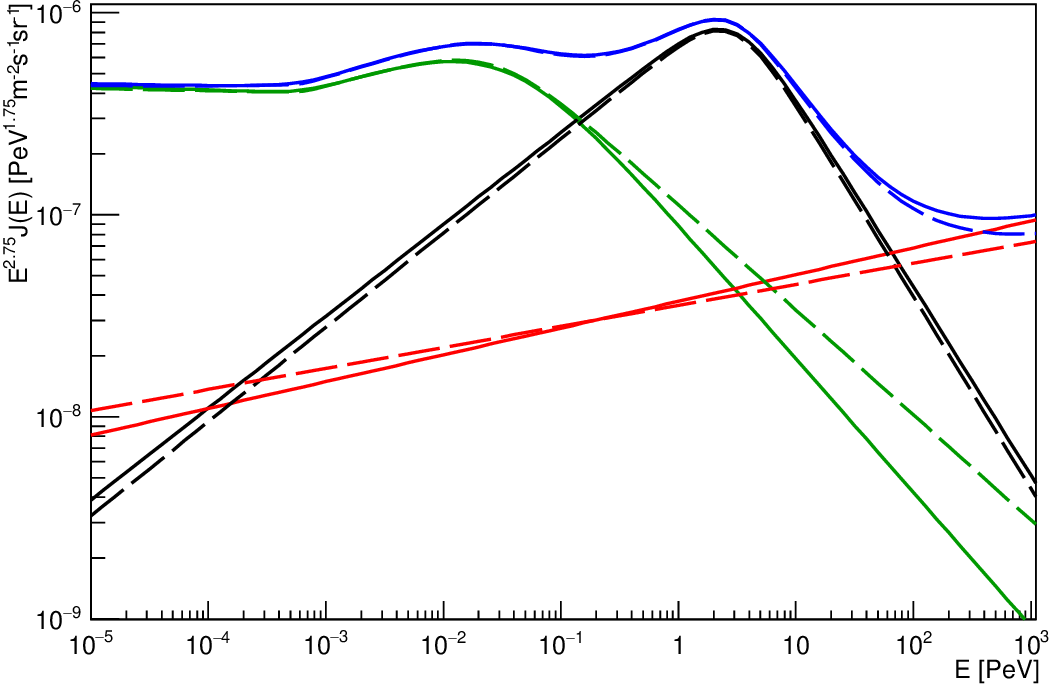}
\caption{Fits to the primary proton spectrum obtained with the MIGRAD (solid curves) and MCMC (dashed curves) methods (see the text for more details). \label{fig:Compare}}
\end{figure}

In Fig.~\ref{fig:Compare} we show the fitting curves for the proton CR spectrum obtained with the MIGRAD and MCMC methods. The first (low-energy) Galactic components are shown as green curves, the second (high-energy) Galactic components --- as black curves, the extragalactic components are shown in red, and the overall fitting curves --- in blue. The difference between the overall fitting curves for the two methods is relatively small at $E < 100$~PeV.

\section{Table of parameters for protons}\label{sect:proton-table}

In Table~\ref{tab:protons} we present the best-fit parameter values for the proton CR spectrum obtained with the MIGRAD (second column) and MCMC (third column) methods. The uncertainties of the parameters estimated with the MCMC method are presented in the last column of the table.

\begin{table*}
\caption{Table of parameters for protons.} \label{tab:protons}
\begin{tabular}{| c | c | c | c c |}
\hline
parameter                                     & value (MIGRAD)       & value (MCMC)           & uncertainties (MCMC)                      &\\
\hline
$K_{1}$ [PeV$^{-1}$m$^{-2}$s$^{-1}$sr$^{-1}$] & 2.05$\times 10^{-4}$ & 2.14$\times 10^{-4}$   & +1.8$\times 10^{-5}$ -2.3$\times 10^{-5}$ &\\
$\gamma_{11}$                                 & 2.769                & 2.761                  & +1.4$\times 10^{-2}$ -9.6$\times 10^{-3}$ &\\
$\gamma_{12}$                                 & 2.571                & 2.590                  & +9.6$\times 10^{-3}$ -1.4$\times 10^{-2}$ &\\
$\gamma_{13}$                                 & 3.413                & 3.270                  & +1.7$\times 10^{-1}$ -1.6$\times 10^{-1}$ &\\
$E_{a}$  [PeV]                                & 6.29$\times 10^{-4}$ & 6.27$\times 10^{-4}$   & +3.3$\times 10^{-5}$ -3.8$\times 10^{-5}$ &\\
$E_{k1}$ [PeV]                                & 3.50$\times 10^{-2}$ & 2.69$\times 10^{-2}$   & +8.2$\times 10^{-3}$ -6.4$\times 10^{-3}$ &\\
$w_{a}$                                       & 3.52                 & 8.12                   & +18.9 -4.4                                &\\
$w_{k1}$                                      & 1.20                 & 1.59                   & +0.41 -0.30                               &\\
\hline
$K_{2}$ [PeV$^{-1}$m$^{-2}$s$^{-1}$sr$^{-1}$] & 1.44$\times 10^{-4}$ & 1.34$\times 10^{-4}$   & +2.8$\times 10^{-5}$ -2.7$\times 10^{-5}$ &\\
$\gamma_{21}$                                 & 2.30                 & 2.28                   & +6.6$\times 10^{-2}$ -6.1$\times 10^{-2}$ &\\
$\gamma_{22}$                                 & 3.68                 & 3.70                   & +0.17 -0.15                               &\\
$E_{k2}$ [PeV]                                & 2.93                 & 2.90                   & +0.34 -0.25                               &\\
$w_{k2}$                                      & 2.44                 & 2.65                   & +0.44 -0.34                               &\\
\hline
$K_{3}$ [PeV$^{-1}$m$^{-2}$s$^{-1}$sr$^{-1}$] & 1.54$\times 10^{-5}$ & 1.58$\times 10^{-5}$   & +9.4$\times 10^{-6}$ -7.6$\times 10^{-6}$ &\\
$\gamma_{3}$                                  & 2.62                 & 2.65                   & +4.9$\times 10^{-2}$ -7.3$\times 10^{-2}$ &\\
\hline
\end{tabular}
\end{table*}

\section{Table of parameters for Helium, CNO and Iron}\label{sect:nuclei-table}

In Table~\ref{tab:nuclei} we present the best-fit parameter values for the Helium (second and third columns), CNO (fourth column), and Iron (last column) nuclei obtained with the MIGRAD method. The two fits for Helium correspond to Fig.~\ref{fig:Helium} and Fig.~\ref{fig:Helium2}, respectively.

\begin{table*}
\caption{Table of parameters for Helium, CNO and Iron.} \label{tab:nuclei}
\begin{tabular}{| c | c | c | c | c c |}
\hline
parameter                                     & Helium (fit 1)       & Helium (fit 2)       & CNO                  & Iron                 &\\
\hline
$K_{1}$ [PeV$^{-1}$m$^{-2}$s$^{-1}$sr$^{-1}$] & 1.54$\times 10^{-4}$ & 1.47$\times 10^{-4}$ & 1.20$\times 10^{-4}$ & 1.36$\times 10^{-4}$ &\\
$\gamma_{11}$                                 & 2.759                & 2.766                & 2.754                & 2.765                &\\
$\gamma_{12}$                                 & 2.355                & 2.353                & 2.400                & 2.580                &\\
$\gamma_{13}$                                 & 2.760                & 3.300                & 3.205                & 3.200                &\\
$E_{a}$  [PeV]                                & 1.50$\times 10^{-3}$ & 1.50$\times 10^{-3}$ & 7.00$\times 10^{-3}$ & 1.14$\times 10^{-2}$ &\\
$E_{k1}$ [PeV]                                & 3.45$\times 10^{-2}$ & 8.24$\times 10^{-2}$ & 0.146                & 0.468                &\\
$w_{a}$                                       & 1.07                 & 1.22                 & 1.94                 & 9.76                 &\\
$w_{k1}$                                      & 4.73                 & 1.46                 & 4.00                 & 3.50                 &\\
\hline
$K_{2}$ [PeV$^{-1}$m$^{-2}$s$^{-1}$sr$^{-1}$] & 4.34$\times 10^{-5}$ & 7.30$\times 10^{-4}$ & 4.35$\times 10^{-5}$ & 2.09$\times 10^{-5}$ &\\
$\gamma_{21}$                                 & 1.91                 & 2.30                 & 2.40                 & 2.30                 &\\
$\gamma_{22}$                                 & 4.18                 & 3.73                 & 3.24                 & 3.71                 &\\
$E_{k2}$ [PeV]                                & 5.92                 & 6.06                 & 30.0                 & 62.9                 &\\
$w_{k2}$                                      & 1.26                 & 3.54                 & 3.00                 & 3.11                 &\\
\hline
\end{tabular}
\end{table*}

\section{Table of parameters for the joint fit}\label{sect:joint-table}

In Table~\ref{tab:joint} we present the best-fit parameter values for the joint fit obtained for protons, the Helium, CNO, and Iron nuclei simultaneously. Two fits were obtained: the first with the MIGRAD method (second column) and the second with the MCMC method (third column). The uncertainties of the parameters for the joint fit estimated with the MCMC method are presented in the last column of Table~\ref{tab:joint}.

\begin{table*}
\caption{Results for the joint fit.} \label{tab:joint}
\begin{tabular}{| c | c | c | c c |}
\hline
parameter                                       & value (MIGRAD)       & value (MCMC)           & uncertainties (MCMC)                      &\\
\hline
$K_{1-p}$ [PeV$^{-1}$m$^{-2}$s$^{-1}$sr$^{-1}$] & 1.96$\times 10^{-4}$ & 1.81$\times 10^{-4}$   & +9.6$\times 10^{-6}$ -9.4$\times 10^{-6}$ &\\
$K_{1-h}$ [PeV$^{-1}$m$^{-2}$s$^{-1}$sr$^{-1}$] & 1.57$\times 10^{-4}$ & 1.43$\times 10^{-4}$   & +7.6$\times 10^{-6}$ -7.4$\times 10^{-6}$ &\\
$K_{1-n}$ [PeV$^{-1}$m$^{-2}$s$^{-1}$sr$^{-1}$] & 1.14$\times 10^{-4}$ & 1.05$\times 10^{-4}$   & +5.2$\times 10^{-6}$ -5.3$\times 10^{-6}$ &\\
$K_{1-f}$ [PeV$^{-1}$m$^{-2}$s$^{-1}$sr$^{-1}$] & 1.34$\times 10^{-4}$ & 1.23$\times 10^{-4}$   & +4.9$\times 10^{-6}$ -5.0$\times 10^{-6}$ &\\
$\gamma_{11}$                                   & 2.765                & 2.773                  & +6.9$\times 10^{-3}$ -6.5$\times 10^{-3}$ &\\
$\Delta \gamma_{21-p}$                          & 0.342                & 0.399                  & +5.6$\times 10^{-2}$ -4.8$\times 10^{-2}$ &\\
$\Delta \gamma_{21-h}$                          & 0.583                & 0.671                  & +6.8$\times 10^{-2}$ -6.1$\times 10^{-2}$ &\\
$\Delta \gamma_{21-n}$                          & 0.519                & 0.606                  & +7.0$\times 10^{-2}$ -6.1$\times 10^{-2}$ &\\
$\Delta \gamma_{21-f}$                          & 0.408                & 0.484                  & +6.8$\times 10^{-2}$ -5.9$\times 10^{-2}$ &\\
$\Delta \gamma_{32}$                            & 1.078                & 1.238                  & +9.9$\times 10^{-2}$ -9.0$\times 10^{-2}$ &\\
$E_{a-p}$  [PeV]                                & 1.07$\times 10^{-3}$ & 1.27$\times 10^{-3}$   & +2.0$\times 10^{-4}$ -1.6$\times 10^{-4}$ &\\
$E_{k1-p}$ [PeV]                                & 2.45$\times 10^{-2}$ & 2.50$\times 10^{-2}$   & +7.0$\times 10^{-4}$ -1.4$\times 10^{-3}$ &\\
$w_{a}$                                         & 1.59                 & 1.42                   & +0.14 -0.12                               &\\
$w_{k1}$                                        & 1.00                 & 1.00                   & +5.2$\times 10^{-2}$ -2.4$\times 10^{-2}$ &\\
\hline
$K_{2-p}$ [PeV$^{-1}$m$^{-2}$s$^{-1}$sr$^{-1}$] & 1.70$\times 10^{-4}$ & 1.98$\times 10^{-4}$   & +8.0$\times 10^{-6}$ -8.5$\times 10^{-6}$ &\\
$K_{2-h}$ [PeV$^{-1}$m$^{-2}$s$^{-1}$sr$^{-1}$] & 7.93$\times 10^{-5}$ & 1.02$\times 10^{-4}$   & +8.1$\times 10^{-6}$ -8.9$\times 10^{-6}$ &\\
$K_{2-n}$ [PeV$^{-1}$m$^{-2}$s$^{-1}$sr$^{-1}$] & 4.50$\times 10^{-5}$ & 5.40$\times 10^{-5}$   & +9.0$\times 10^{-6}$ -8.2$\times 10^{-6}$ &\\
$K_{2-f}$ [PeV$^{-1}$m$^{-2}$s$^{-1}$sr$^{-1}$] & 2.63$\times 10^{-5}$ & 3.58$\times 10^{-5}$   & +3.9$\times 10^{-6}$ -4.0$\times 10^{-6}$ &\\
$\gamma_{2}$                                    & 2.37                 & 2.42                   & +0.15 -0.16                               &\\
$\Delta \gamma_{2}$                             & 1.37                 & 1.15                   & +0.09 -0.08                               &\\
$E_{k2-p}$ [PeV]                                & 3.13                 & 2.88                   & +0.17 -0.17                               &\\
$w_{k2}$                                        & 2.99                 & 3.69                   & +0.55 -0.41                               &\\
\hline
$K_{3-p}$ [PeV$^{-1}$m$^{-2}$s$^{-1}$sr$^{-1}$] & 1.92$\times 10^{-5}$ & 1.00$\times 10^{-5}$   & +3.9$\times 10^{-6}$ -2.7$\times 10^{-6}$ &\\
$\gamma_{3-p}$                                  & 2.62                 & 2.57                   & +4.4$\times 10^{-2}$ -4.5$\times 10^{-2}$ &\\
\hline
\end{tabular}
\end{table*}

\section{Comparison of joint fits obtained with different methods}\label{sect:comparison-all}

\begin{figure*}
\vspace{0.2cm}
\includegraphics[width=\textwidth]{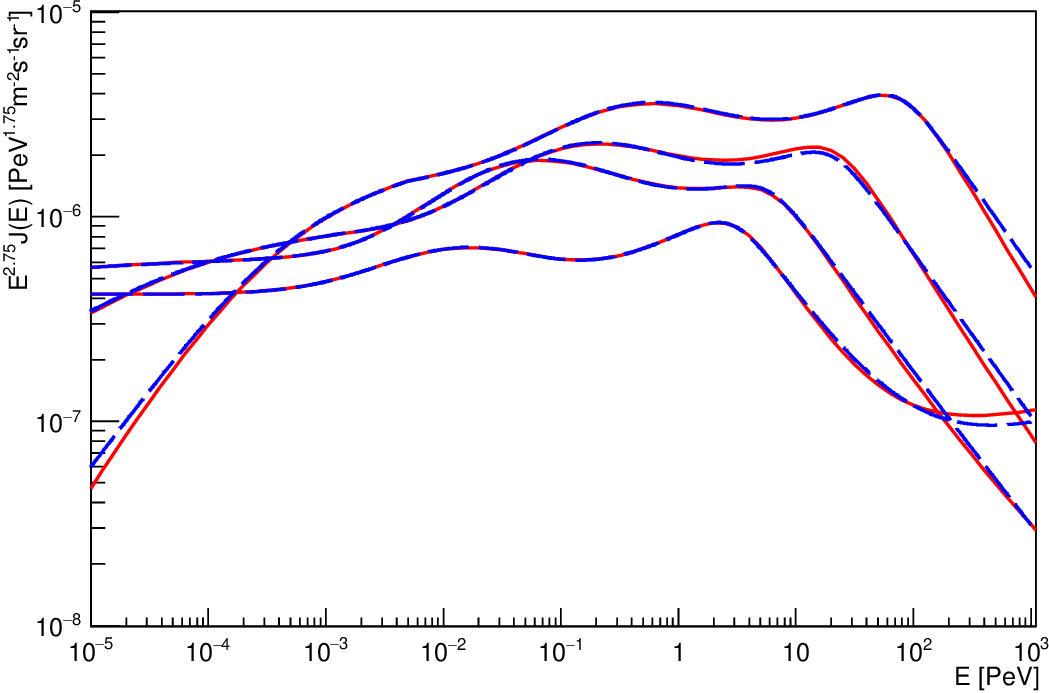}
\caption{Joint fits to the primary proton, Helium, CNO, and Iron spectra obtained with the MIGRAD (solid red curves) and MCMC (dashed blue curves) methods (see the text for more details). \label{fig:Compare-All}}
\end{figure*}

In Fig.~\ref{fig:Compare-All} we show the fitting curves for the proton, Helium, CNO, and Iron elemental spectra. These fitting curves are results from the joint fit (see Appendix~\ref{sect:joint-table}). For better visibility we have multiplied the spectra by a factor of 2$^{n_{v}}$ with $n_{v} = 0, 1, 2, 3$ for the protons, Helium, CNO, and Iron nuclei, respectively. The difference between the overall fitting curves for the two methods is relatively small at $E = 100$~GeV--$100$~PeV.

\bsp	
\label{lastpage}
\end{document}